# Sequence to Sequence Neural Speech Synthesis with Prosody Modification Capabilities

*Slava Shechtman*[1], *Alex Sorin*[1]

[1]IBM Research, Haifa, Israel

slava@il.ibm.com, sorin@il.ibm.com

## Abstract

Modern sequence to sequence neural TTS systems provide close to natural speech quality. Such systems usually comprise a network converting linguistic/phonetic features sequence to an acoustic features sequence, cascaded with a neural vocoder. The generated speech prosody (i.e. phoneme durations, pitch and loudness) is implicitly present in the acoustic features, being mixed with spectral information. Although the speech sounds natural, its prosody realization is randomly chosen and cannot be easily altered. The prosody control becomes an even more difficult task if no prosodic labeling is present in the training data. Recently, much progress has been achieved in unsupervised speaking style learning and generation, however human inspection is still required after the training for discovery and interpretation of the speaking styles learned by the system.

In this work we introduce a fully automatic method that makes the system aware of the prosody and enables sentence-wise speaking pace and expressiveness control on a continuous scale. While being useful by itself in many applications, the proposed prosody control can also improve the overall quality and expressiveness of the synthesized speech, as demonstrated by subjective listening evaluations. We also propose a novel augmented attention mechanism, that facilitates better pace control sensitivity and faster attention convergence.

## 1. Introduction

Nowadays, sequence to sequence neural TTS systems can generate speech with close-to-natural speech quality [1][2]. These systems have as an input a linguistic sequence (e.g. characters or phonemes, enriched with other text-based features, such as lexical stress, phrasing, word embedding, etc.) with an optional speaker label (for multi-speaker settings) and output a speech acoustic sequence, represented by frame-wise spectral parameters (e.g. mel-spectrum), from which a waveform can be generated using a neural vocoder [3][4]. They generate speech prosody (phone durations, pitch and loudness) implicitly, so their prosody control is not straightforward. Hence, while the output speech is generally natural-sounding, the prosody tends to be rather unpredictable, if not guided.

Alternative neural TTS systems [5][6] predict the speech acoustics from the symbolic sequence augmented with detailed prosodic information (including at least phone durations and pitch trajectory). In such systems, the speech prosody is trained and predicted by a separate network and fed into the speech acoustics prediction network. While this enables fully controllable prosody, the quality of the synthesized speech is somewhat worse, compared to the jointly trained seq2seq neural TTS systems [1][2].

In many applications, such as chat bots, virtual assistants, virtual sales agents, etc., there is a requirement to control the TTS prosody (e.g. speaking style, speaking rate, expressiveness) at inference time, based on directives, coming from outside of the TTS system. The state-of-the-art approaches for adding such functionality to the seq2seq neural TTS are detailed below:

1. (Semi-)Supervised approaches, that exploit prosodic and/or speaking style labeling (partially or fully generated by human subjects) [7]. Those approaches are straightforward, but the human labeling is expensive, error prone and time consuming. Very few labeled open source resources exist for speech synthesis.

2. Exemplar-based prosody control. The acoustic/prosodic realization of speech can be transferred from a given spoken example (uttered by any speaker), using appropriate latent space representation [8]. Although this approach can result in an impressive quality [8], its use case is special and not feasible in most real-life TTS applications.

3. Unsupervised speaking style learning approaches. The speech acoustics' latent space is automatically trained so that the latent dimensions are disentangled to enable their independent manipulation at inference time [9]. However, human inspection is still required after the training for the speaking style discovery and interpretation.

In this work we propose an alternative approach, in which the neural sequence to sequence TTS is made aware of prosody through a so-called *prosody info*, i.e. a small set of naturally disentangled and interpretable prosodic observations (e.g. distinct duration and pitch statistics observations), evaluated at the utterance level. During the training, the system is exposed to the measured *prosody info* vectors, while at inference time the *prosody info* vectors are predicted from the linguistic input. Furthermore, they can be externally modified, component by component, to deliberately shift prosodic characteristics of the synthesized speech (e.g. change speech pace, pitch variability, etc.)

The proposed neural TTS system follows Tacotron2 architecture [2], comprising a recurrent attentive sequence to sequence network for spectral feature prediction, cascaded with Wavenet[3]-like neural vocoder [10], with several modifications, intended to improve the synthesized speech quality, training convergence and sensitivity to the proposed prosody control mechanism. Specifically, to improve the prosody control sensitivity, we propose a novel *augmented attention* mechanism that facilitates better pace control sensitivity and faster attention convergence.

The paper is structured as follows. In Section 2 the spectrum feature prediction network is reviewed, elaborating on several minor modifications of the original Tacotron2 system [2], while the novel augmented attention mechanism is presented in the dedicated Section 3. In Section 4 the proposed prosody control mechanism is described. Section 5 describes the Wavenet-like neural vocoder [10] modifications. Following that, Section 5 presents experimental results, demonstrating, among others, that the proposed prosody control can improve overall quality and expressiveness of the synthesized speech.

## 2. The mel-spectra prediction network

The mel-spectral feature prediction module follows the Tacotron2 [2] architecture, comprising the convolutional encoder with a terminal recurrent layer (bidirectional [11] LSTM [12]), converting the symbolic sequence to the sequence of their latent representations, cascaded with the autoregressive attentive decoder that expands the encoded symbolic sequence to a sequence of fixed-frame mel-spectral feature vectors.

The Tacotron2 decoder [2] predicts one spectral frame at a time from the pre-net-processed previous spectral frame, $s_p$, conditioned on the input context vector, $x_c$, generated by the attention module. The decoder generates its hidden state vector, $h_c$, with two-layered stacked LSTM network. The hidden state vector $h_c$, combined with the input context vector $x_c$, is fed to the final linear layers to produce the current mel-spectrum and the end-of-sequence flag. At the end, there is also a convolutional post-net that refines the whole utterance mel-spectrogram to improve its fidelity.

The original Tacotron2 model can consume text characters directly [2], but to simplify the training, we feed the system with the sequence of symbols from an extended phonetic dictionary, comprising the phone identity, lexical stress (3-way: primary, secondary, unstressed) and phrase type (4-way: affirmative, interrogative, exclamatory, other), enriched with distinct word break and silence symbols. This symbolic input sequence is generated by an external grapheme to phoneme rule-based TTS Front End module, adopted from the unit selection TTS [13].

### 2.1. Differential spectral loss

We experimentally found that a slightly better synthesized speech quality is obtained when incorporating the mean square error (MSE) applied to the difference between the current and the previous mel-spectra into the final system loss [2].

Let $y_t$ be the predicted mel-spectrum at time $t$ before the post-net [2], $z_t$ be the final predicted mel-spectrum at time $t$ and $q_t$ be the mel-spectrum target at time $t$. Then, the spectral loss is given by:

$$Loss_{spc} = 0.5 MSE(y_t, q_t) + \\ 0.25 MSE(z_t, q_t) + 0.25 MSE(z_t - z_{t-1}, q_t - q_{t-1}) \quad (1)$$

### 2.2. Double feed training

In Tacotron2 the training procedure follows the teacher-forcing approach, i.e. the prediction of the current mel-spectrum is performed based the *real* previous mel-spectrum, processed by the pre-net, as opposed to the inference procedure, where the prediction is autoregressive [2]. In our system we applied a *double feed* during the training, i.e. we feed the decoder's pre-net with both the true previous mel-spectrum and the predicted one, concatenated together. At inference time, when the true frames are not available, the predicted mel-spectrum is simply duplicated. While increasing the total network size by only 0.1%, this modification reduces the total model regression loss by about 15%, as tested on two professionally recorded US English speech corpora of 13 and 22 hours.

## 3. Augmented attention mechanism

The attention mechanism is essential in the decoder module for automatic alignment between the decoder input and output sequences. At each output time step $t \epsilon \mathbb{Z}, 0 \leq t < T$, corresponding to the output frame sequence index $t$, the vector of attention weights is produced (i.e. *alignment vector* $a_t[n], 0 \leq n < N$) determining the relative attention (i.e. the *alignment* probabilities) of the $t$-th output to the whole input sequence at its indices $n \epsilon \mathbb{Z}, 0 \leq n < N$. Once the alignment vector is determined, the input context vector $x_c$ is obtained as a linear (convex) combination of the decoder input sequence vectors with the alignment vector serving as its corresponding weights. The input context vector is stored as a part of the decoder state and fed to the next decoder stages to finalize the output sequence prediction. This process is repeated until the end-of-sequence symbol is generated.

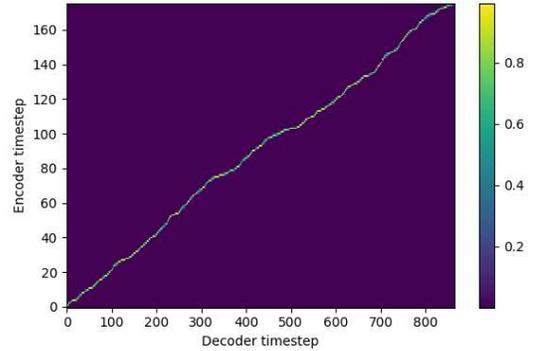

Figure 1: *Alignment matrix visualization.*

As presented on Figure 1, The $N \times T$ matrix of the attention probabilities (a.k.a. the *alignment matrix*, having the alignment vectors as its columns) represents the input-output sequence alignment, that directly determines the speaking pace and strongly influences the speaking style of the synthesized audio. In addition, speech synthesis quality and intelligibility heavily depend on the alignment matrix structure (e.g. monotonicity, sparseness, unimodality, etc.), and it should be preserved when controlling the speech generation process at inference time.

The original Tacotron2 system [2] deploys the additive attention with content-based and location-based mechanisms, where the latter is fed with both the previous alignment vector and the cumulative alignment vector (summed up to the previous time step $t$-1).

When experimenting with the prosody controls, concatenated to the decoder input sequence at each time step (see Section 4), we noticed that the original attention mechanism is not responsive enough to the pace modification directives. In addition, the prosody control occasionally resulted in local or global loss of attention, expressed in local or global quality degradations, occasional phone misses or early stoppings.

To improve the system robustness and responsiveness, we came up with an *augmented attention* mechanism, that can be applied above any existing attention mechanism within a recurrent decoder, as denoted in Figure 2.

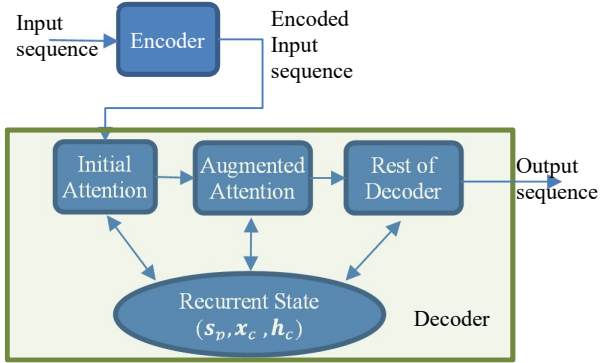

Figure 2: *Sequence to sequence prediction network with augmented attention*

The basic idea of the proposed augmented attention is as follows. Being aware of the desired alignment matrix structure (e.g. monotonic alignment evolvement), we first derive a set of feasible alignment vector candidates from the previous alignment vectors, in addition to the initial current alignment vector. Then, we apply a soft-selection to obtain the final alignment vector in such a way that its expected structure (i.e., a unimodal shape with a sharp peak) is preserved.

Let $b_\tau[n]$ be the initial alignment vector, as evaluated by the initial attention module and $a_\tau$ be a final alignment vector at output time step $\tau$. Assuming the monotonic attention [14], without skipping input symbols, we create a candidate set $\{c_k\}_{k=0}^2$ by adding the previous alignment vector $b_{t-1}[n]$ together with its shifted version $b_{t-1}[n-1]$ to the current initial alignment $b_t$ at the current time step $t$:

$$\{c_k\}_{k=0}^2 = \{b_t, b_{t-1}[n], b_{t-1}[n-1]\} \quad (2)$$

This augmentation assumes that at the current time we either stay aligned to the previous input symbol or move to the next one.

Having this set of candidates, one can just train a soft-selection to come up with the final alignment vector, but to be on the safe side, we want to ensure that the soft-selector prefers properly structured candidates to eliminate occasional attention corruption.

To that end, we propose a scalar structure metric $f(c)$ that assesses the unimodality and the peak sharpness of the alignment vector candidate $c$. This metric is differentiable and confined in $[0,1]$ interval. It combines a *LogSumExp* soft maximum assessment $f_1(c)$ with additional *peak sharpness* metric $f_2(c)$ to ensure that non-sharp and/or non-unimodal alignment vectors attain low scores. Here $f_2(c)$ is derived from the common "peakiness" metric [15], i.e. L2-norm divided by L1-norm, by noting that L1-norm is always unity for the alignment vectors and the squared L2-norm of the worst-case flat alignment vector equals to 1/N. The boost constant 1.67 was experimentally set to reduce the sensitivity of this metric.

The combined structure metric we propose is given by:

$$f(c) = \lfloor f_1(c) f_2(c) \rfloor_{0.12} \quad (3)$$

where

$$f_1(c) = 0.1 \cdot \log \sum_{n=0}^{N-1} \exp(10 \cdot c[n]), \quad (4)$$

$$f_2(c) = \min\left(1.67 \cdot \left(\frac{N \cdot \|c\|_2^2}{(N-1)} - \frac{1}{N-1}\right), 1\right), \quad (5)$$

and the thresholding operator $\lfloor x \rfloor_\alpha$ is defined as:

$$\lfloor x \rfloor_\alpha \equiv \begin{cases} x, & x > \alpha \\ 0, & else \end{cases}$$

The added thresholding operation, with the experimentally set near-zero threshold of 0.12, comes to ensure that bad alignment vector candidates do not fit for the soft-selection.

The structure-preserving soft-selection of the alignment vector is performed in two stages. The first stage is given by:

$$d = \alpha\, b_{t-1}[n-1] + (1-\alpha)\, b_{t-1}[n], \quad (6)$$

where $\alpha$ is a scalar initial stage selection weight, generated by a single fully connected layer, fed with the concatenated decoder state variables $(s_p, x_c, h_c)$ and terminated with the sigmoid layer. Observing the first stage selection (6) one can note that it provides an explicit phone transition control by the embedded prosody info, which is a part of the input context vector (see Section 4 for details).

The final stage of the selection procedure utilizes the structure metric $f(c)$:

$$a_t = (1-\gamma)\beta d + \gamma(1-\beta)b_t, \quad (7)$$

where $\beta$ is a scalar final stage selection weight, generated by a single fully connected layer, fed with the input context vector $x_c$ and terminated with the sigmoid layer, and $\gamma = f(b_t)(1 - f(d))$ is a structure preference score. This multiplicative structure preference score ensures that the initial attention vector will be considered only if its structure is preferable over the other candidate.

The first stage of the proposed augmented attention mechanism formulation resembles to some extent the *forward attention with transition agent* [16], that also functions as a post-processing of an initial attention mechanism. However, a product of attention vectors in the core of the forward attention algorithm [16] encourages occasional loss of attention. On the other hand, the proposed augmented attention is additive and structure-preserving, so the final attention vector's structure is guaranteed to be at least not worse than the initial one's.

The augmented attention does not only improve the prosody control responsiveness, but also improves the alignment convergence during the training, as can be seen in Figure 2, showing the average alignment vector entropy of the validation set (100 sentences) during the training of a data corpus of 13K sentences.

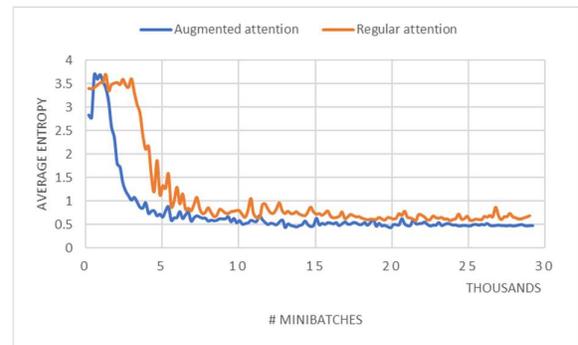

Figure 3: *Average entropy of alignment vectors of the validation set during the training (minibatch size=48)*

## 4. Prosody control

In human speech the same verbal information can be conveyed in many ways. The encoder output sequence (see Figure 4) encapsulates all the verbal information used in the system, while the proposed *prosody info* observations, extracted from the recordings during the training, provide additional hints on how to convey that verbal information. We want prosody info components to be disentangled and easily interpretable, e.g. having distinct components for pace, pitch and loudness. In this work we used only two components (pace and pitch control) for the prosody info (omitting the loudness control, as our voice corpora had uniform loudness):

1. The *log* of the average phoneme duration (excluding silences) as a pace measure of an utterance;

2. The span of *log*-pitch evaluated as 0.95-quantile minus 0.05-quantile of an utterance log-pitch trajectory.

The prosody info vectors are normalized per speaker, component-wise, so that their efficient span ($median \pm 3std$) is mapped to [-1,1].

The prosody info vector is embedded into a 2-dimensional latent space (by a single fully connected unbiased layer with *tanh* nonlinearity) and concatenated with each vector in the encoder output sequence. Consequently, the decoder is exposed to the prosody info through the input context vector.

During the training we feed the observed prosody info vector to the system. To ease the alignment convergence at initial steps of the training, we set the prosody info to zero for the first 5 epochs (about 1500 mini-batch steps in our case).

After the training is complete, we train a prosody-info prediction module separately, minimizing MSE loss. The prosody module is fed with the encoder output sequence and predicts the prosody info out of it. The prediction is done with 3-layered stacked LSTM (128 cells at each layer), followed by a linear layer that produces the prosody info vector (with the output size of 2).

At inference time (see Figure 4), the prosody info is predicted from the encoder output sequence and can be deliberately changed by adding a component-wise offset (in $[-1,1]$ range).

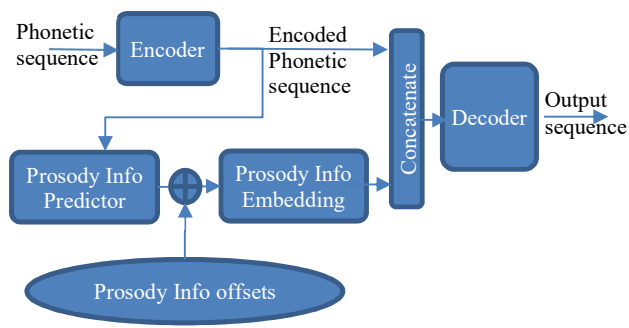

Figure 4: *Proposed prosody control at inference time.*

We experimented with the proposed prosody info control on several voice corpora and found that the system is responsive to component-wise prosody info inference-time modifications, successfully slowing down/speeding up as a response to the pace component modification and increasing/decreasing expressivity as a response to the pitch component modification.

## 5. Neural vocoder

The output of the spectral feature prediction network is fed to a neural vocoder [2][4][10] to generate the waveform. We use a Wavenet-style vocoder [2] which is based on the CUDA reference implementation[1] of the classic autoregressive Wavenet architecture [3]. This architecture is composed from a stack of dilated convolutional layers with residual connections followed by a softmax layer for prediction of the audio sample probability distribution. High-frequency resolution mel-spectrograms are used for local conditioning. To this end the frame-wise mel-spectra are upsampled to the audio sampling rate using a transposed convolution layer, go through $1 \times 1$ convolutional layers and concatenated with the activations of the dilated convolution layers.

Our modifications to the reference implementation[1] are described below.

The reference implementation[1] is tuned for 16 kHz audio sampling rate. To support 22 kHz sampling rate, we modified the mel-spectrum frame hop length from 200 to 256 samples and the analysis window size from 800 to 1024. The mel-filter bank size remains 80 channels.

Like in [10], we performed the nearest mel-spectra picking for the upsampling. We also increased the number of the residual connection channels from 64 to 128.

The convnet-based Wavenet operates on audio quantized using 256-levels µ-law encoding. It's well known that the µ-law quantization and Wavenet prediction errors are transformed to an audible wide-band noise in the synthesized audio. Several noise-shaping methods have been proposed to alleviate this problem. We use a first order pre-emphasis filter as a noise shaping means proposed in [4]. The vocoder is trained to predict pre-emphasized mu-law quantized audio, while de-emphasis is applied to the predicted audio at inference time. Thus, the neural vocoder accepts the mel-spectrograms originated from the original audio (without pre-emphasis). We found that this setup is favorable, because in the case of the pre-emphasized mel-spectra the attention weights of the preceding mel-spectra prediction network are not properly learnt.

Although this type of noise-shaping greatly improves the speech quality, the de-emphasis sometimes introduces local audible saturation effects in the synthesized signal. We use an adaptive gain control (AGC) procedure to avoid the saturation. This procedure introduces a small look ahead which does not prevent the synthesized audio streaming in an application that requires it. The de-emphasized audio samples are first calculated without the dynamic range limiting. Then, the saturation occurrences are detected, and a smooth gain curve is calculated to ensure that audio fits in the 16-bit dynamic range.

While experimenting with this neural vocoder, we found out that audible artefacts in the generated signals sometimes occurred at speech onsets that follow silent parts of the audio signal. The most probable explanation of this effect is that the network is trained with randomly extracted one-second-long audio segments that frequently start with high energy (e.g. the segment can start in the middle of a vowel) and learns to predict the energy bursts from silent initial conditions. To overcome this problem, we train only with the audio segments beginning at silent audio regions. The silent regions are automatically detected based on a short-time energy analysis.

---

[1] https://github.com/NVIDIA/nv-wavenet

We assessed the modified neural vocoder performance in the analysis-synthesis task by a subjective crowd MOS evaluation, in which the samples generated by this vocoder were compared to the natural speech. The neural vocoder was trained on male and female high-quality US English speaker audio data of 13 and 10 hours correspondingly. The MOS figures obtained in this analysis-synthesis evaluation are provided in Table 1. The degradation for the female voice was statistically non-significant

Table 1. *Neural vocoder analysis-synthesis MOS (μ±95%)*

| MOS | PCM | Vocoder |
|---|---|---|
| Female | 4.09±0.05 | 4.00±0.05 |
| Male | 4.37±0.04 | 4.28±0.04 |

# 6. Experiments

In our experiments we used male and female proprietary voice corpora with sampling rate of 22050 Hz, that were originally created for building a product level concatenative TTS system. The male dataset contains 13 hours of speech and the female dataset contains 22 hours of speech. Both were produced by native US English speakers and recorded in a professional studio. The audio was recorded utterance by utterance, where most of utterances contain a single sentence. For both voices we assessed the feasibility of the proposed prosody control technique. A subset of samples for the experiments described below is accessible online[1].

First, we performed informal, qualitative listening, for which we generated grid tables for component-wise prosody info modifications (±0.1, ±0.5) for several samples per voice, compared to the samples with unmodified *prosody info*. Our impression was that the prosody control modified the samples as expected. The pace component makes the synthesized speech faster or slower while the pitch component modifies the pitch variability and thus influences the perceived expressiveness of the synthesized speech. We did not encounter any intelligibility or early stopping problems.

Second, we conducted two formal MOS listening test experiments (one per voice corpus), in which we assessed several systems' quality and expressiveness. We used each one of the systems to synthesize a set of 40 held-out sentences and evaluated them together with the original held-out recordings of the same voice. All the samples were normalized to the same energy. The tests were performed using the Amazon Mechanical Turk (AMT) platform with 90–140 anonymous and untrained subjects participating in several evaluation sessions, constructed so that each sentence is evaluated by 25 distinct subjects. The subjects listened to one utterance at a time and evaluated its quality and expressiveness separately at a five-point categorical scale ("Poor", "Bad", "Fair", "Good", "Excellent") and the corresponding queries ("Rate the overall quality and naturalness…", "Rate the expressiveness…"). The scores for each system were calculated as the average over all its sentences. The systems that participated in each test included:

- "PCM": held out speech recordings;
- "WORLD": WORLD [17] based neural TTS at 22KHz as described in [18];
- "RegAttn": the proposed seq2seq system with the regular attention mechanism and no prosody control;
- "AugAttn(0,0)": the proposed seq2seq system with the augmented attention mechanism and the predicted prosody info (no prosody info modification);
- "AugAttn(*a*, *b*)": the proposed seq2seq system with the augmented attention mechanism and the predicted prosody info, with additive offsets *a* and *b* for the pace and the pitch components correspondingly.

In Table 2 and Table 3 the evaluation results for the female and the male voice are presented. The significance analysis for the results in Table 2 revealed that most of cross-system expressiveness differences are statistically significant, except of the difference between *RegAttn* and *AugAttn*(0,0) and the difference between *RegAttn*(-0.1,0.5) and *AugAttn*(0.15,0.6). In terms of naturalness, all the augmented attention systems performed like *RegAttn* (no significant difference), except of *RegAttn*(-0.1,0.5) that performed slightly better (*p*=0.046). So, for the female voice the prosody control was able to significantly improve the perceived expressiveness, while preserving the original quality and naturalness.

Similarly, the significance analysis for the male voice (Table 3) revealed that only the pair *RegAttn* and *AugAttn*(0,0), as well as the pair *RegAttn*(0.2,0.8) and *AugAttn*(0.5,1.5) are equivalent in terms of perceived expressiveness. In terms of naturalness, both *RegAttn*(0.2,0.8) and *AugAttn*(0.5,1.5) provide significant improvement, compared to *RegAttn*(0,0) and *RefAttn*. So, for the male voice the prosody control was able to significantly improve expressiveness, quality and naturalness altogether.

Table 2. *MOS Evaluation for Naturalness and Expressiveness for US English female voice (μ±95%)*

| MOS | PCM | WORLD | RegAttn | AugAttn(*pace*, *pitch*) | | |
|---|---|---|---|---|---|---|
| | | | | (0,0) | (-0.1,0.5) | (0.15,0.6) |
| Nat. | 4.17 ± 0.06 | 3.00 ± 0.09 | 3.86 ± 0.06 | 3.86 ± 0.06 | 3.93 ± 0.06 | 3.93 ± 0.06 |
| Expr. | 4.27 ± 0.05 | 3.22 ± 0.08 | 3.88 ± 0.06 | 3.83 ± 0.06 | 4.04 ± 0.06 | 4.01 ± 0.06 |

Table 3. *MOS Evaluation for Naturalness and Expressiveness for US English male voice (μ±95%)*

| MOS | PCM | WORLD | RegAttn | AugAttn(*pace*, *pitch*) | | |
|---|---|---|---|---|---|---|
| | | | | (0,0) | (0.2,0.8) | (0.5,1.5) |
| Nat. | 4.09 ± 0.05 | 2.58 ± 0.07 | 3.68 ± 0.06 | 3.71 ± 0.05 | 3.81 ± 0.05 | 3.81 ± 0.05 |
| Expr. | 4.20 ± 0.05 | 2.97 ± 0.06 | 3.76 ± 0.05 | 3.74 ± 0.05 | 3.88 ± 0.05 | 3.93 ± 0.05 |

To conclude, the experiments revealed that the proposed controllable sequence to sequence neural TTS is responsive to the prosody controls at inference time, while preserving high quality and naturalness. We also showed that the prosody

---

[1] http://ibm.biz/Bdz2Jm

controls can be applied to improve perceived expressiveness of the voice.

## 7. Summary

In this work we introduced a sequence to sequence neural TTS system with interpretable and disentangled prosody controls of speaking pace and expressiveness. The proposed prosody control mechanism was demonstrated to improve expressiveness of the synthesized speech, while preserving or improving its quality and naturalness. Several modifications for Tacotron2-like architecture were proposed, including a novel *augmented attention* mechanism, that improved the robustness and responsiveness of the system to the proposed prosody control mechanism.